# Detailed study of non-equilibrium characteristics of quasi-neutral TNSA plasmas

**Zhe Zhu[1a], A. Bonasera[1,2*], D. Batani[3], M. R. D. Rodrigues[1], K. Batani[4], J.A. Pérez-Hernández[5], M. Ehret[5], E. Filippov[5], H. Larreur[3,6,7], D. Molloy[7,8], G. G. Rapisarda[2,9], D. Lattuada[2,10], G. L. Guardo[2], C. Verona[11], Fe. Consoli[2], G. Petringa[2], A. McNamee[8], M. La Cognata[2], S. Palmerini[12,13], R. De Angelis[13], G. A. P. Cirrone[2], V. Istokskaia[15], R. Lera[5], L. Volpe[5,16], D. Giulietti[17], S. Agarwal[18,19], M. Krupka[18,19], S. Singh[18,19], Jun Xu[20,a]**

[1] Cyclotron Institute, Texas A&M University, College Station, TX, USA

[2] Laboratori Nazionali del Sud, Istituto Nazionale di Fisica Nucleare (LNS-INFN), Catania, Italy

[3] Université de Bordeaux, 33405 Talence, France

[4] IPPLM Institute of Plasma Physics and Laser Microfusion, Warsaw, Poland

[5] CLPU (Centro de Láseres Pulsados), Villamayor, Spain

[6] Universidad de Salamanca, Salamanca, Spain

[7] HB11 Energy Holdings Pty, Freshwater, NSW, Australia

[8] Queen's University Belfast, School of Mathematics and Physics, Belfast, UK

[9] Dipartimento di Fisica e Astronomia "E. Majorana", Università di Catania, Catania, Italy.

[10] Facoltà di Ingegneria e Architettura, Università degli Studi di Enna "Kore", Enna, Italy

[11] Dipartimento di Ingegneria Industriale, Università di Roma "Tor Vergata", Roma, Italy

[12] Dipartimento di Fisica e Geologia, Università degli Studi di Perugia, Perugia, Italy

[13] Istituto Nazionale di Fisica Nucleare, sezione di Perugia, Perugia, Italy

[14] ENEA - Frascati, Fusion and Nuclear Safety Department, Via E. Fermi 45, 00044, Frascati, Italy.

[15] ELI Beamlines Facility, The Extreme Light Infrastructure ERIC, Dolni Brezany, Czech Republic

[16] ETSIA, Universidad Politécnica de Madrid, Madrid, Spain

[17] Dipartimento Fisica, "E. Fermi", Università di Pisa and INFN, Pisa, Italy

[18] FZU-Institute of Physics of Czech Academy of Sciences, Prague, Czech Republic

[19] Faculty of Mathematics and Physics, Charles University, Prague, Czech Republic.

[20] School of Physics Science and Engineering, Tongji University, Shanghai 200092, China.

[a] also at Shanghai Institute of Applied Physics, CAS, Shanghai 201800, China

## Correspondence:

* A. Bonasera: abonasera@comp.tamu.edu, D. Batani: dimitri.batani@u-bordeaux.fr

## Abstract

In an experiment performed in November 2022 at the petawatt (PW) laser facility at Vega III located in Salamanca-Spain, we have studied the successful production of several radioisotopes using protons accelerated by the Target Normal Sheath Acceleration (TNSA) mechanism (Rodrigues et al. [1]). The experimental proton energy distribution recorded on a shot-to-shot basis and confirmed in a follow up experiment (K. Batani et al. [2]), allowed us to derive the number of nuclear reactions taking place in different targets on a single shot. From this analysis, using the ratio of the yields $^{11}C/^{7}Be$, we obtained an effective "single shot" temperature of the TNSA plasma. We used this value to evaluate the yield of $\alpha$ particles from the reaction $p+^{11}B\rightarrow 3\alpha$ which may reach $1.6\pm0.5\times10^{9}$ $\alpha$ in $2\pi$. From the fluctuations of the protons and the fusion yields, we derived a "TNSA-Equation of State" (EoS), The deviation of such "EoS" from the classical ideal gas limit is well reproduced by the soliton solution of the Korteweg-de Vries (KdV) equation for each shot.

## I. Introduction.

The Chirped Pulse Amplification (CPA) [3] technique allowed the production of petawatt lasers at high repetition rates. Such high-intensity lasers are currently used in a variety of basic-physics experiments as well as to develop important applications with societal impact, including medical applications as demonstrated by different experimental results from several groups [1,2,4,5,6].
When a high-intensity laser beam impinges on a thin target, it produces a beam of energetic protons reaching tens of MeV, a mechanism commonly referred as Target Sheath Normal Acceleration (TNSA) [7- 10]. The accelerated protons can be used to produce nuclear reactions on a secondary target [4-6].
We have used this approach in an experiment performed at the Centro de Laseres Pulsados (CLPU) in Salamanca, Spain, in November 2022 [1,7, 8]. The main goal of the experiment was to try to show the possibility of using lasers to produce radioisotopes of medical interest (see refs. [1] and [2]). In particular we used targets containing natural boron (which is 20% $^{10}B$ and 80% $^{11}B$) to produce ions like $^{7}B$ and $^{11}C$. The production of radioisotopes can in turn be used to characterize the laser-accelerated protons.
In this paper we analyze the data from that experiment (as well from a follow up experiment [2]) to obtain two main results:
1) From the ratio of the yields $^{11}C/^{7}Be$, we obtain an effective temperature T of the TNSA plasma, i.e. the ensemble of protons and electrons originating from the contaminant layer present on the rear side of the Al target and accelerated in vacuum.
2) We show that the experimentally measured proton energy per particle vs. T deviates from the classical ideal gas limit. Such deviations are well reproduced by using the soliton solution of the KdV equation for each shot.

In the experiment at CLPU, the laser beam energy was kept near the nominal maximum value of 30 J (roughly corresponding to 7J on target after passage in the compressor and in focusing optics) with ~200 fs pulse duration (for a complete description of the experimental set-up we refer to [1,7,8]). The produced protons were recorded on a shot-to-shot basis using Micro Channel Plates located at the back of a Thompson Spectrometer (MCPTS) which is an innovation respect to the usual method of adopting Image Plates (IPTS) detectors [11-13]. It is well known that IP are insensitive to the strong electronic disturbance due to laser matter interaction (EMP-Electro Magnetic Pulse), but usually they can only provide protons and charged ion energy distributions averaged over shots. Even when the proton flux is enough to record a spectrum on IP (which is usually happening when laser with much bigger energies are sued), still using IPTS on a single shot basis would require opening the interaction chamber after each shot. The MCPTS was localized at zero degrees respect to the normal to the Al target which produced energetic protons and other ions (from impurities) [1]. However, in a follow up experiment performed at the same facility in March 2023 [2,11,12], the MCPTS was replaced by a IPTS. In this case, many shots realized in the same nominal conditions were recorded on the same IPTS, with the same laser and target parameters used in the first experiment. In both experiments, ion distributions were also determined on a shot-to-shot basis at different angles using diamond detectors in Time of Flight (TOF) configuration. The different detectors showed large fluctuations from shot to shot in ion spectra. In this work we will take advantage of this feature, first, to deduce possible fluctuations on the nuclear reactions, and second, to derive the "TNSA-EoS".

The paper is organized as follows. In section II, we obtain the nuclear reaction rates from the proton distribution for a single shot and the average reaction rates. Both the single shot and the average proton energy distributions display an exponential decrease at high energies which can be fitted using Maxwellian distributions for each case. The temperatures entering the thermal distributions are independently determined by the ratios of radio isotopes yields measured experimentally [1, 2, 11, 12]. In section III, an "effective plasma temperature" is derived for each shot from the ratio of $^{11}C/^{7}Be$ produced in $p+^{nat}B$ reactions. Such "effective temperature" is the temperature that a thermalized plasma should have to produce the same nuclear yield ratios. We show that the corresponding Maxwellian proton distributions agree rather well with the measured distributions for each shot and on average. The deviation from the ideal gas case gives important information on the TNSA-EoS and on the role of solitons in the plasma dynamics as discussed in section IV. In section V, we compare our results to previous experimental data and highlight the role of solitons. Conclusions and suggestions for future experiments are drawn in section VI. Finally, a brief presentation of the experimental layout and a comparison of the average proton and electron distributions recorded at different angles are given in appendix A.

## II. Shot to shot derivation of nuclear reaction rates.

In Fig. 1, we plot the proton energy distributions obtained using the MCPTS and the IPTS detectors. On the left panel (Fig. 1A), the MCPTS distribution is averaged over 43 shots (red triangles) [1,12] and the IPTS distribution is averaged over 20 shots (blue circles) [2, 11]. In the plot, we also included a Maxwellian distribution (dashed line) discussed in detail in the sequel.

In ref. [1] we have used the MCPTS average distribution and compared to the phenomenological distribution of ref. [6] to predict the number of nuclear reactions for different channels. A reasonable agreement to the experimental data was obtained [1]. In Fig. 1 (B), we plot the distributions for the individual shots. Fluctuations are very large especially at energies around 1 MeV; this may have strong impact on radioisotopes production for nuclear reactions with negative Q values. Following ref. [14], the detailed knowledge of the proton energy distribution function, the range of the protons in the target (from SRIM [15] or other models), and the reaction cross sections (or equivalently the astrophysical S-factors) [15-16] allow us to study the nuclear dynamics in the plasma [18].

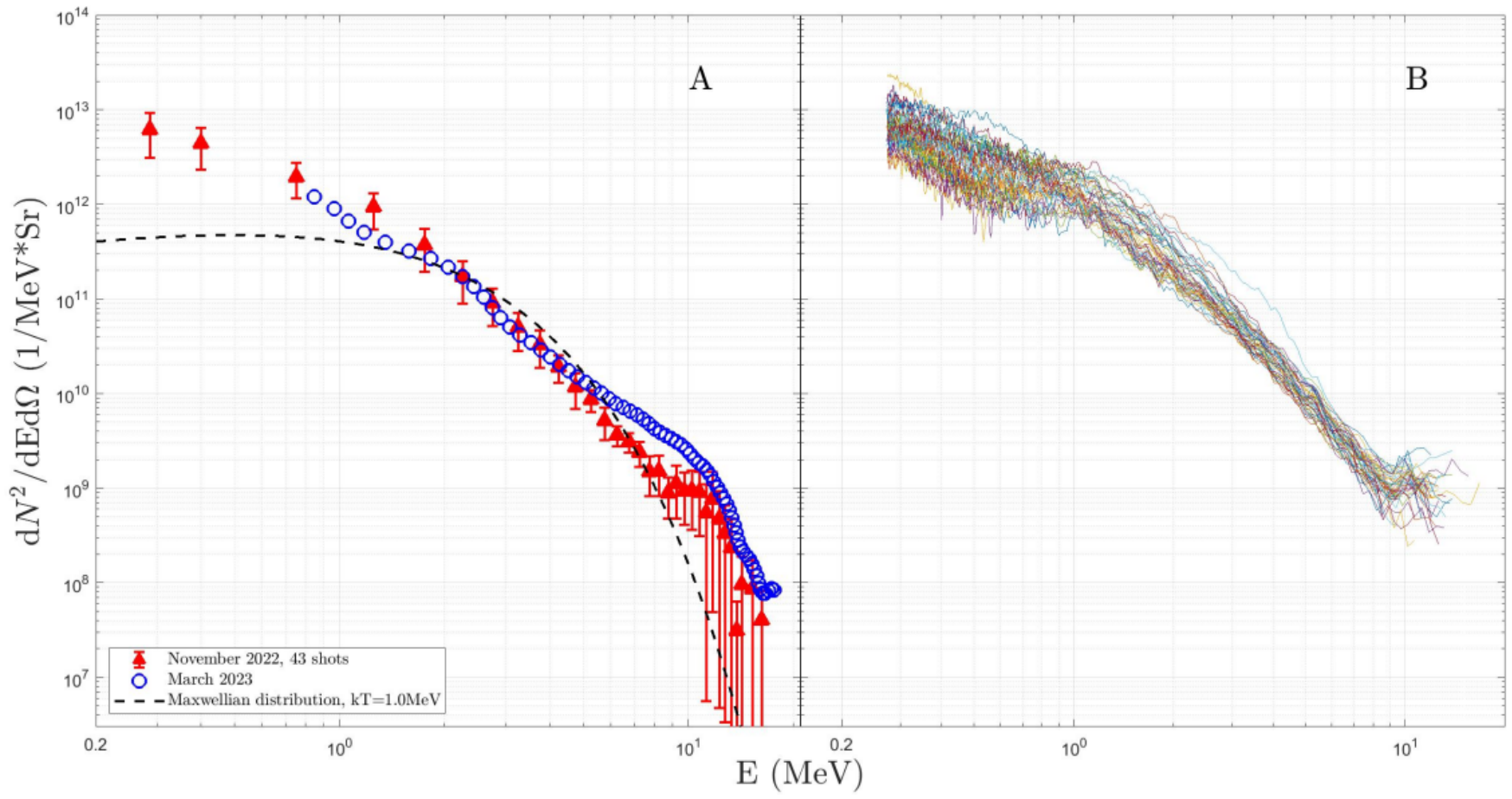

Figure 1. A) Proton energy distributions recorded using a IPTS in March 2023 compared to the average (over shots) recorded by a MCPTS detector in November 2022. A Maxwellian distribution with an "effective" temperature T=1.0 MeV is also displayed (see text). B) Shot to shot proton energy distributions recorded with the MCPTS, error bars (statistical) are not included for clarity of presentation.

In this work we modify the approach and use the radio-isotopes average reaction rates from [1], to derive, shot-to-shot, the reaction rate for each isotope using the proton energy distribution functions plotted in Fig.1 (B). We discuss the following reactions:

1) $^{63}$Cu(p,n)$^{63}$Zn, Q = -4.15 MeV;
2) $^{10}$B(p,$\alpha$)$^{7}$Be, Q = 1.14 MeV;
3) $^{11}$B(p,n)$^{11}$C, Q = -2.76 MeV;
4) $^{11}$B(p,$\alpha$)2$\alpha$, Q = 8.6 MeV.

Experimental reaction results are reported in our previous work [1], apart for the 3$\alpha$ reaction for which we will here derive constraints on its production rate from the ratio of $^{11}$C/$^{7}$Be yields [17, 18, 19]. Reaction (1) was investigated by locating natural Cu targets with different thicknesses at different angles in the forward and back direction with respect to the laser beam. All the other reactions were investigated in the so-called pitcher-catcher scheme, i.e. by locating a thick natural B target about 2cm away from the Al thin target used for TNSA acceleration of protons and ions. In this case protons distributions cannot be measured at zero degrees; thus, we assume that their distributions are the same as those measured by the TP when no catcher is used. Repeating the shots under similar experimental conditions (and even in the two different experimental campaigns) supports this assumption.
Reaction rates were obtained from the isotope decays measured with a High Purity Germanium Detector (HPGE) [1,2]. The data were collected accumulating a certain number of shots, and the decays were measured as function of time. From this quantity and knowing the decay constant $\lambda$ of the isotope we can derive an average number of decays $N_0$ using the relation [1]:

$$N(t) = \sum_{i=0}^{n} N_{0i} e^{-\lambda(t-t_{shot\,i})} \approx N_0 \sum_{i=0}^{n} e^{-\lambda(t-t_{shot\,i})} \qquad (1).$$

Where $t_{shot\ i}$, is the time when shot i occurs, and n is the number of shots. We can improve eq. (1) since the proton energy distribution is known for each shot (see Fig. 1B). If $dN_0/d\Omega$ is the experimental value [1] for the average number of isotopes produced, for a single shot we define:

$$\frac{dN_i}{d\Omega} = \frac{n\int_{E_{th}}^{Emax} \frac{dN_{pi}}{dEd\Omega}\sigma(E)R(E)\left(1-e^{-\frac{d}{R(E)}}\right)dE}{\sum_j^n \int_{E_{th}}^{Emax} \frac{dN_{pj}}{dEd\Omega}\sigma(E)R(E)\left(1-e^{-\frac{d}{R(E)}}\right)dE}\frac{dN_0}{d\Omega} \qquad (2).$$

In eq. 2 (see also eqs. 2 and 3 in ref. [1]), $\frac{dN_{pi}}{dEd\Omega}$ is the experimentally measured proton energy (E) distribution for the i-shot given in Fig.1 (B), σ(E) is the reaction cross section for the isotope of interest parametrized from experiments [1], d is the thickness of the material where the energetic ions propagate and R(E) is the proton range from SRIM [15] (or equivalent models). $E_{th}$ is the lowest proton energy needed to produce the isotope, equal to -Q for reactions with negative values of Q, and zero for reactions with positive ones. With respect to the approximation used in eq. 1, here we assign to each shot a probability proportional to the reaction rates given by the ratio of the integrals. In the ratio, any dependence on the target density, the normalization of the proton energy distribution function, and the range (for thin targets only) cancels out [1]. Finally, to get integrated yields, we multiply the proton energy distribution for an average collection solid angle ΔΩ [1].

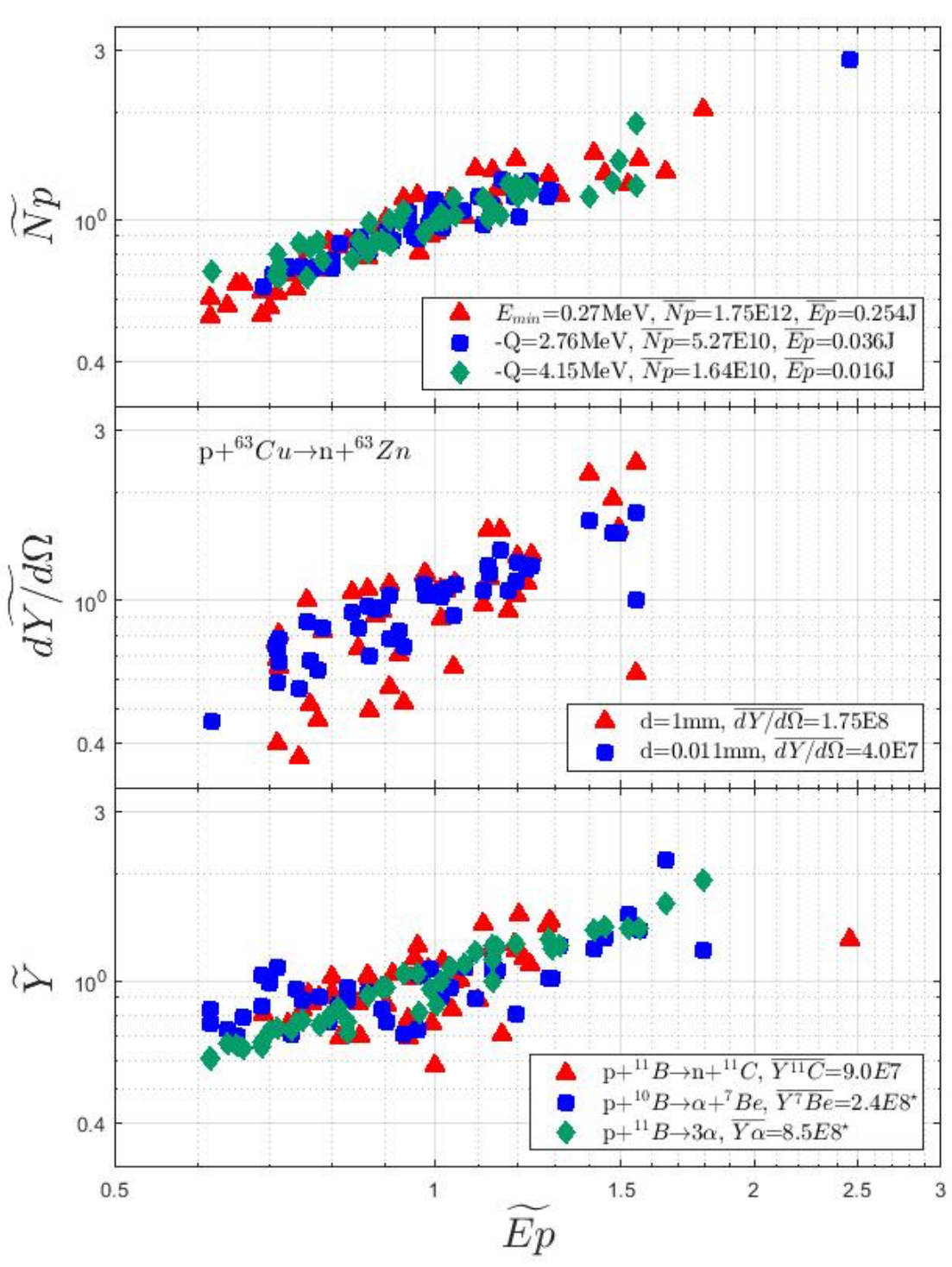

Figure 2. Normalized proton number (top panel), $^{63}Zn$ yield (middle panel) at zero degrees [1] ($dY/d\Omega$), $^{11}C$,$^{7}Be$ yields and 3α integrated over 2π (bottom panel) ($Y^{11}C$, $Y^{7}Be$). In all plots the x-axis represents the normalized proton kinetic energy measured by the MCPTS. The average values for each case are given in the insets and have been corrected for the target concentration. The asterisk in the bottom figure refers to values estimated from the average proton (or $^{11}C/^{7}Be$) ratios, see text.

In Fig. 2, we plot the number of protons vs. their total kinetic energy divided by their respective average values reported in the inset, top panel. The average quantities are obtained by integrating the moments of the experimental proton energy distributions (i.e. times $E^0$ or times $E^1$). Results are plotted for different $E_{th}$ cutoffs given by the value of Q for the reactions 1-4) and the lowest measured proton energy, $E_{min}$. These large fluctuations reflect on the production of $^{63}Zn$, middle panel in Fig. 2, which has the largest negative Q among the reactions discussed in this work. For this case we show the results for two different thicknesses of the Cu target [1]. Notice that the d=0.011 mm target was in the back direction respect to the laser propagation [1]. Thus, in principle, the proton energy distribution and its fluctuations may be different from the measured one and displayed in Fig. 1. However, in ref. [1] it was noted that the proton energy distribution in the back and in the front respect to the Al target give similar nuclear reaction rates.

The bottom part of Fig. 2 shows the yield fluctuations for the reactions of protons with natural B target (corrected by the concentrations). It is evident that fluctuations decrease for positive reactions with positive values of Q and that for some shots we get yields much larger than the average values.

## III. Single shot plasma temperature from nuclear reaction yield ratios.

In Fig. 2, we included the 3α yield which was not discussed in ref. [1]. The difficulty for this channel is due to the discrimination of α from the overwhelming yield of protons and other ions [7,8]. Furthermore, the B targets are rather thick thus most of the produced α remain trapped inside due to teur vey shrt penetration tranges. In ref. [1] a phenomenological estimate of the yield in the conditions of this experiment was given. Now we follow the method of refs. [17, 18, 19] and obtain the α-yield compatible with the reaction rates of the other two reactions occurring in the same target, i.e. $^{11}C$ and $^{7}Be$ [1,2]. We start from the ratio of the reaction yields for two different channels A and B:

$$\frac{Y_A}{Y_B} = \frac{\rho_{(A)} \int_{-Q_A}^{Emax(A)} 2\sqrt{\frac{E}{\pi}}\left(\frac{1}{kT}\right)^{\frac{3}{2}} e^{-\frac{E}{kT}} \sigma_A(E) R_A(E)\left(1-e^{\frac{-d_A}{R_A(E)}}\right) dE}{\rho_{(B)} \int_{-Q_B}^{Emax(B)} 2\sqrt{\frac{E}{\pi}}\left(\frac{1}{kT}\right)^{\frac{3}{2}} e^{-\frac{E}{kT}} \sigma_B(E) R_B(E)\left(1-e^{\frac{-d_B}{R_B(E)}}\right) dE} \qquad (3).$$

$\rho_{(A,B)}$ are the target densities and all the other symbols have already been defined above. In eq. 3 we assume the proton energy distribution to be given by a Maxwellian with "effective" temperature T. We stress that the term "effective" refers to an ideal thermal plasma at temperature T producing the same yield ratios as in the experiment.

The ratio given by eq. 3 is plotted in Fig. 3 (full red line) as function of temperature, for the case $^{11}C/^{7}Be$ measured in ref. [1] (open circles). The experimental value equal to 0.37 univocally determines the effective value T=0.97 MeV. The resulting Maxwellian is plotted in Fig.1 (A) and compared to the measured proton energy distribution function after some (irrelevant) normalization. The Maxwellian reproduces *rather* well the data in the region between 1.0 to 10.0 MeV, which is the region of interest for the reactions at hand. This agrees with ref. [19] where it was shown that the temperature derived from nuclear reaction rates was the same as the one measured from the ion spectra using Faraday cups. In ref. [17] the measured ratio $\alpha/^{7}Be$ was used to obtain the plasma temperature. In the same plot, we include the ratios from eq. 3 for the two other channel combinations.

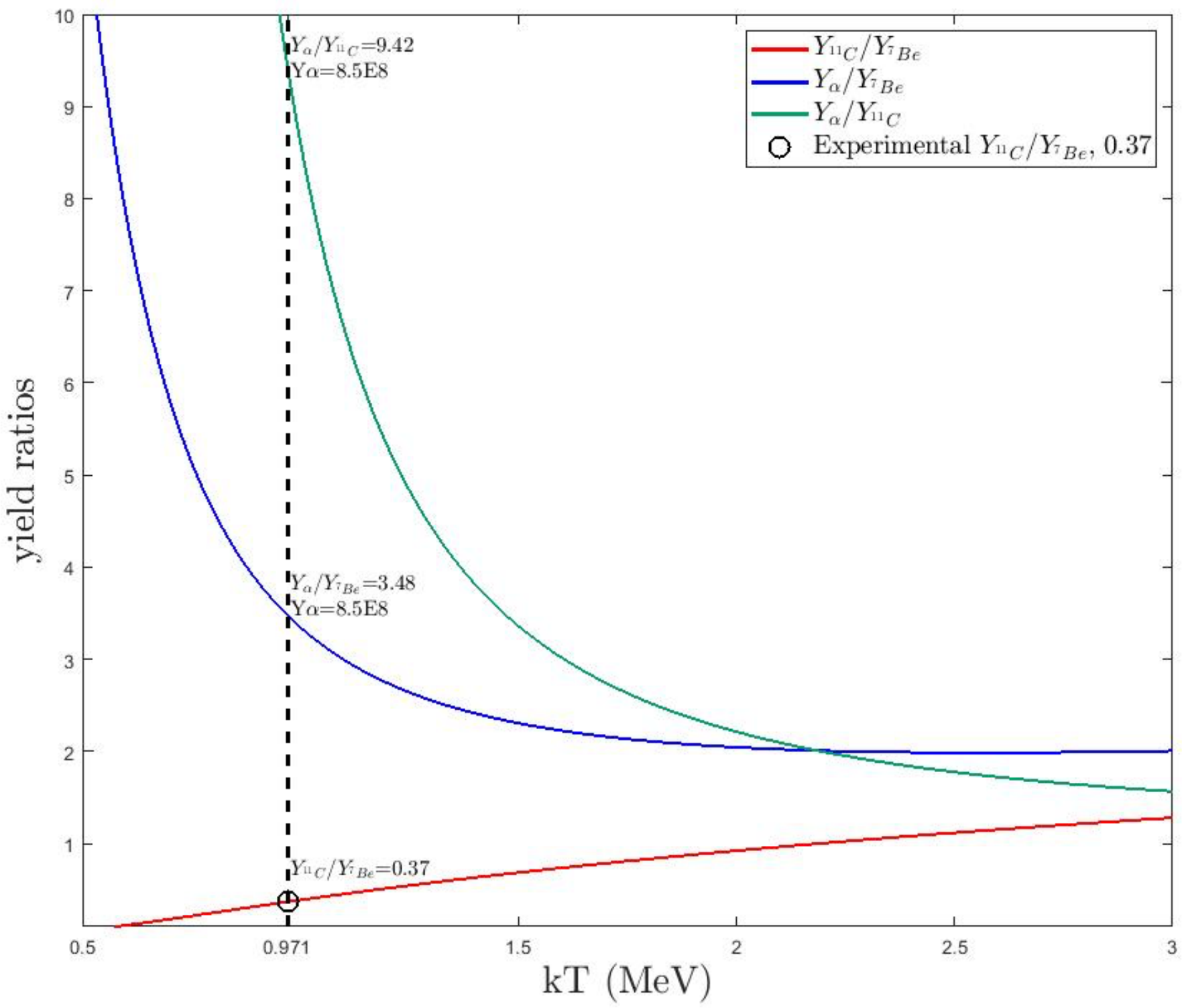

Figure 3. Yield ratios from different pB reactions. The experimental value for $^{11}C/^{7}Be$ is used to determine the effective temperature T~1.0 (0.971) MeV. For this temperature we can determine the ratio for the other reactions and in turn the average yield of α-particles.

From the effective temperature, we can determine the value of the ratios for $\alpha/^{11}C$ and $\alpha/^{7}Be$, which in turn give consistently the same average number of produced α: $8.5\times10^{8}$. Recall that the number of reactions is a factor of 3 less than the number of α, thus it is in good agreement with the estimate of ref.s [1,2,11,12]. Screening effects on the cross sections entering eq. 3 are small at our temperatures furthermore, if present, they should cancel in the ratios since the compound nuclei in the reactions are approximately the same leading to a similar screening potential [16]. The effects induced by range variations could be more important but, in our pitcher-catcher scheme, the catcher is in a "normal" solid state and here the SRIM calculations are reliable since they have been tested in beam target experiments over decades.

It is important to stress the fact that the ratio of the reaction rates $\alpha/^{7}Be$ is of the order of one (3.48/3 where the denominator comes from the number of α produced from reaction 4) above). Recall that these estimates are obtained after correcting for isotope concentration (80% of $^{11}B$ and 20% of $^{10}B$ in natural boron). This implies that the low energy resonances for reaction 4) and the high energy ones for reaction 2) do not play any major role since the proton energies are so high. In practice, the important quantity is the geometrical cross section which is about the same for the two target nuclei since they differ of one mass unit only. Thus, if we want to increase the yield from reaction 4), we need an increase of the low energies (below 1 MeV) proton yield. This observation confirms the estimation of the 3α yield, since such a value must be close to the $^{7}Be$ (+α) production which was directly measured with the HPGE detector in [1,2].

From the knowledge of the average number of α, we can estimate its fluctuations as discussed above. Alternatively, we can repeat the procedure above, i.e. for each shot ratio $^{11}C/^{7}Be$, we can determine the effective temperature and the α-yields. For our follow up discussion, the first approach will be adopted but the other method gives consistent results. From the effective temperature for each shot we can fit a Maxwellian distribution to each experimental proton yield of Fig.1 (B) (not included for clarity of presentation) obtaining similar results to the average case reported in Fig.1 (A).

The α-reaction yields are shown in Fig. 2 (bottom panel) with diamond symbols. The maximum value is $1.6\pm05\times10^{9}$ in 2π since protons were collected from the B target on the front only. This value (and the variance) was obtained averaging the results from the two different methods. The maximum value for the $^{7}Be$ case is $5.2\times10^{8}$ in 2π, close to the 3α channel obtained above. The largest yields may be obtained in different shots. In fact, it is the combination of the proton yields for each energy and their respective ranges (which is higher for high energy protons) which determines the final reaction rates. We expect another factor of two in the same experimental conditions, by locating another B target on the back direction respect to the TNSA Al target, as we discussed for the $^{63}Zn$ case [1]. It is important to stress that the different ratios give consistent scaling (see Fig. 2), even though the three different channels have rather different Q-values.

## IV. Deviations from the ideal gas equation of state.

In Fig. 2, we have plotted the energy, and the number of protons derived from the detector for each shot. We have also derived an effective temperature from the ratio of two competing channels in section III.

In this section, we want to explore the relation connecting the energy per particle (average proton energy) and the temperature in the plasma produced as a result of TNSA acceleration. We describe such relation using the term *Equation of State* (EoS) of the system. This is different from the usual description of EoS as a relation which connects pressure (P), Volume (V) and temperature (T) in an system at equilibrium, however if the density is known, the pressure can be derived from the energy per particle. In this work we may assume that thermal equilibrium is reached locally and at different time steps, since the system is dynamical and not confined [20,21,22].

For a perfect gas the relation connecting the energy per particle and the temperature is simply E/N=3/2 kT. Since, we are dealing with proton temperatures in the MeV region (electron temperatures are similar, see appendix A) and quite low densities, the EoS should in principle be very close to the EoS of a perfect gas (in particular, we do not expect quantum effects to play any role and there is no need of using Fermi-Dirac statistics since the temperature of the system is order of magnitude larger than its Fermi energy). Therefore, we are particularly interested in investigating the departures from the relation between energy per particle and temperature which is valid for a perfect gas. However, we should modify the relation E/N=3/2 kT to consider that in different reactions we are exploring proton energies above a certain energy threshold, $E_{th}$. Thus, using a Maxwellian at temperature T, we can estimate numerically the average energy per particle:

$$\langle E\rangle = \frac{E_p}{N_p} = \frac{\int_{E_{th}}^{\infty} 2\sqrt{\frac{E}{\pi}}\left(\frac{1}{kT}\right)^{\frac{3}{2}} e^{-\frac{E}{kT}}\, E\, dE}{\int_{E_{th}}^{\infty} 2\sqrt{\frac{E}{\pi}}\left(\frac{1}{kT}\right)^{\frac{3}{2}} e^{-\frac{E}{kT}}\, dE} \qquad (4),$$

and similarly using the experimental proton distribution displayed in Fig.1 (B) for each shot. Since the distribution contain an exponential factor the main contribution to the calculation of $\langle E \rangle$ comes from low energies, close to $E_{th}$, therefore obviously $\langle E \rangle > E_{th}$ but they are of the same order of magnitude.

In Fig. 4, we plot the average proton energy obtained from the MCPTS vs. the effective temperature calculated from the $^{11}C/^{7}Be$ ratio shot-to-shot (see Fig. 3). The 0.27 MeV threshold energy corresponds roughly to the lowest sensitivity of the detector, while for the other cases to -Q for $^{11}C$ and $^{63}Zn$ (for reference). The ideal gas limit with such cutoffs, eq. 4, is given by the open symbols and reproduces the behavior of the TNSA plasma within a factor of 2. No error bars are reported but from the errors on the average ratio values and the error on the proton energy distribution, we estimated at most a 20% error for each point.

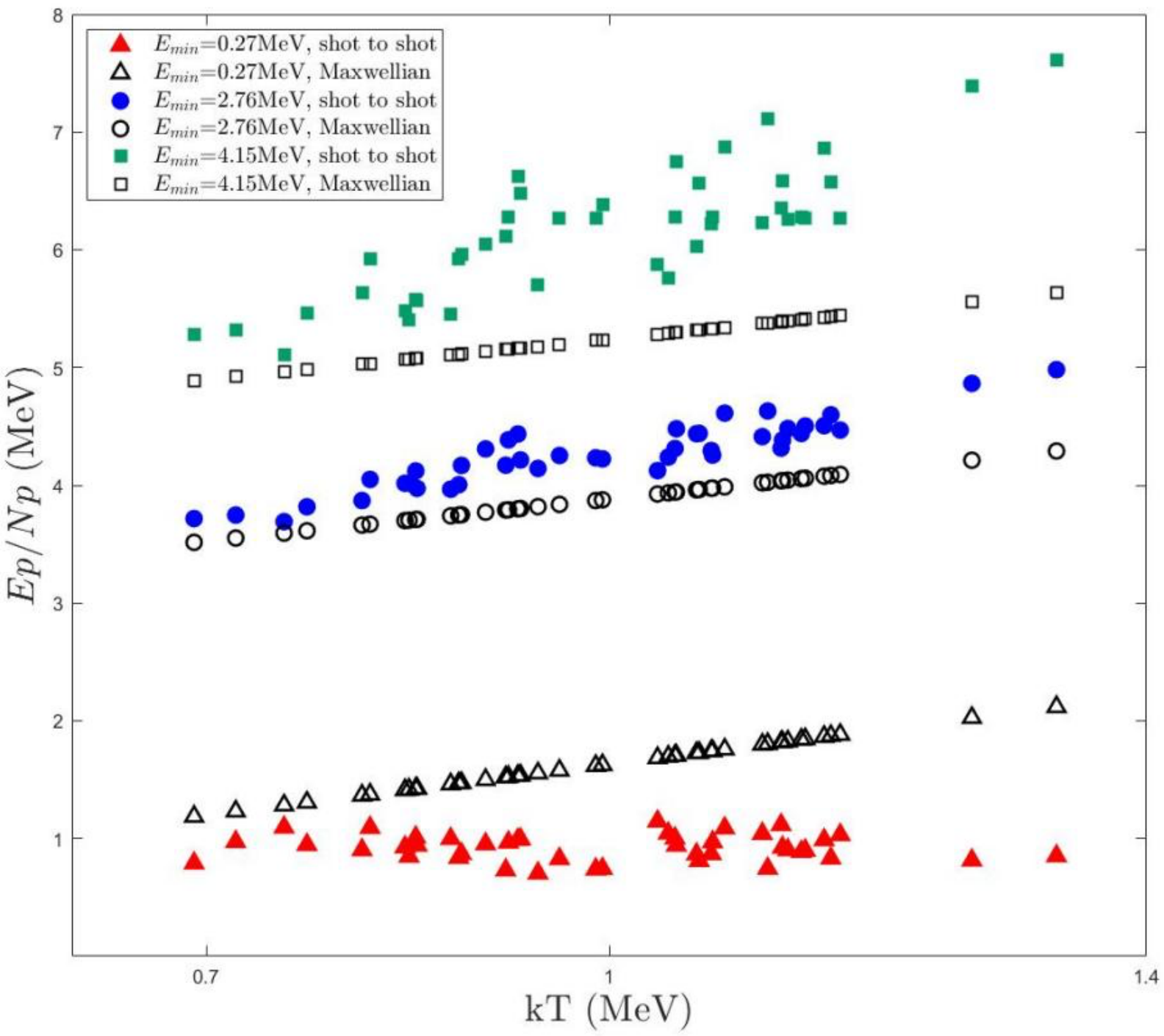

Figure 4. Average proton energy for different cutoff energies vs. shot-to-shot temperatures obtained from the $^{11}C/^{7}Be$ ratios. Data are given by the full symbols while the open symbols refer to the ideal gas limit given by eq. 4 in the text. Temperature derived from the ratios of the two other competing channels give consistent results.

It is important to notice in Fig. 4 that the proton energy per particles is above the ideal gas limit for large cutoff energy (squares), and below it for low cutoff energy (triangles).

In TNSA proton acceleration, the electrons are relativistic because of their much smaller mass and leave the plasma earlier than the protons and other heavier ions. This creates a charge separation and a Coulomb field which is responsible for proton acceleration (deceleration) above (below) the classical ideal gas limit. The deviations from the ideal gas limit can then be evaluated by the difference δE between the average proton energy obtained from the experiment (Fig. 1B) and the one calculated from eq. 4 for a Boltzmann distribution. This corresponds to the difference between the full and the open symbols in Fig. 4 for each temperature.

In the TNSA accelerated plasma, the electrons and the ion tend to move with different velocities due to their difference in mass, thus creating *charge separation waves* which generate strong transient fields (as experimentally evidenced in several papers). Such transient electric fields can be the origin of the differences $\delta E$.

The *charge separation waves* correspond to transient local perturbations of the plasma neutrality which is a characteristic of an ideal plasma. Such perturbations of neutrality are expected to be small at each time steps since, essentially, the transient electric fields in the plasma act to remove the excess charge to restore plasma neutrality. Hence the TNSA plasma is at all times close to neutrality, i.e. quasi-neutral.

The characteristic time of plasma evolution can be estimated from

$$t = \frac{\frac{1}{2}d}{\sqrt{\frac{2E_{th}}{m_p}}} = \frac{d}{2v} \qquad (5).$$

where d=6μm is the thickness of the TNSA target used in our experiment and $E_{th}$ is taken form eq. 4. This is nothing else than the average time needed by a proton of mass $m_p$= 938 MeV and energy $E_{th}$ to cross the Al target. Short times imply high cutoff energies and vice versa. The time $\tau^*$ needed by the plasma to reach neutrality is obtained when the measured proton distribution and the ideal gas limit coincide, or $\delta E(t^*)=0$ (see circle symbols in Fig. 4). In practice, the time $\tau^*$ corresponds to the time needed by protons moving with the velocity corresponding to energy $E_{th}$ (close to $\sim\langle E\rangle$) over the distance of reference in our problem (~d/2).

In Fig. 5, we plot $t^*$ vs. the laser duration time $\tau_L$. The results show that the time for the plasma to reach neutrality is of the same order of magnitude of the laser duration but, within the range measured in our experiment, almost independent on it. The time $\tau^*$ allows defining a typical velocity $v_s = \frac{d}{2\tau^*} = \sim 2\ 10^9$ cm/s for 6 μm thickness. This is $\sim 6.10^{-2}$ of the velocity of light c and gives an estimate of the speed associated to the propagation of the perturbation.

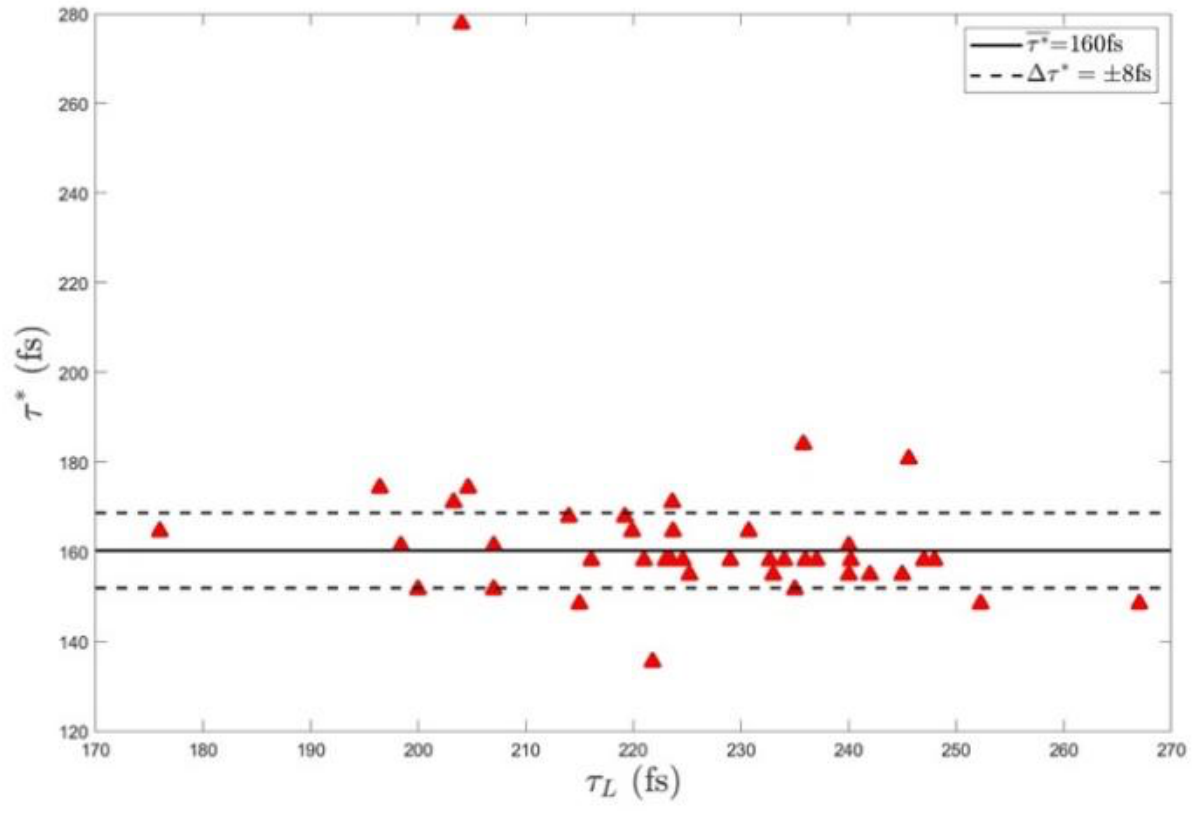

Figure 5. Time $\tau^*$ to reach neutrality vs laser duration time.

A quasi-neutral plasma, as discussed above, supports the propagation of solitary waves solutions, i.e. solitons, waves which propagate without change of shape. Here we will show that that the perturbations of plasma neutrality can be described using KdV equation, hence they have a solitonic character, and we will show that this conclusion is supported by the data discussed in this work. Following Lifshits-Piatyevski [23] we assume a dependence of the perturbations on $\xi = z - ut$, corresponding to solitons propagating with speed $u$, along the z-direction. The time evolution of the soliton velocity field $a(\xi)$ is derived in ref. [23,24]. Here we write it as:

$$a_{\tilde{t}} + a a_{\tilde{\xi}} + \beta a_{\widetilde{\xi\xi\xi}} = 0. \qquad (6).$$

The subscripts indicate derivative respect to time $\tilde{t} = \frac{t}{\tau^*}$ and to the variable $\tilde{\xi} = \tilde{z} - u\tilde{t}$, where $\tilde{z} = \frac{2z}{d}$ is the distance in the laser direction, and all quantities are dimensionless. This equation was originally derived to describe solitons in shallow water by Korteweg-de Vries (KdV) [23]. The solution of eq. 6 is well known to be:

$$a = a_1 \left[ cosh^{-2} \left( \frac{\tilde{z} - u\tilde{t}}{2} \sqrt{\frac{a_1}{3\beta}} \right) \right] \qquad (7).$$

In eq. 7, $a_1$ (the soliton amplitude), u, and β are free parameters that may be fitted to data. The relation between the soliton velocity distribution and the potential field is given by [21]:

$$\left( \frac{e}{m_p} \right) \delta E = \frac{1}{2} a^2 - \frac{1}{2} (a - v)^2 \qquad (8).$$

In our case $\delta E$ refers the difference between the measured proton energy and the ideal gas limit as defined in relation to Fig. 4. In eq. 8, e=1 is the proton charge, and v its velocity which is much smaller than electron velocity which is of the order of c (electrons being relativistic). Since the position $\tilde{z}$ is not measured in the experiment, we will estimate it by defining $\tilde{z} = u\widetilde{t_M}$; the parameter $\widetilde{t_M}$ will be derived below. The second condition is that $\delta E(\tilde{t} = 1) = 0$, which gives $a(\tilde{t} = 1) = \frac{v(\tilde{t}=1)}{2}$ from eq. 8. The latter equation can be used to fix the parameter $a_1$ in eq. 7:

$$a_1 = \frac{v(\tilde{t}=1)}{2} cosh^2[\eta(\tilde{t}_M - 1)]; \qquad \eta = \frac{u}{2} \sqrt{\frac{a_1}{3\beta}} \qquad (9).$$

where all free parameters have been included in one parameter η which depends on $a_1$ as well.

Eq. 8 has a very transparent physical meaning: the field deviation is due to the motion of the protons in the plasma which creates some charge unbalance [8,10,25,26]. When the proton velocity v=0, the plasma is neutral, δE=0 and the ideal gas EoS is recovered. We can solve eq. 8 to derive the wave propagation speed $a(\xi \rightarrow \tilde{t})$. Simple algebra gives:

$$a(\tilde{t}) = \frac{\tilde{t}}{v_s} \left[ \frac{\delta E}{m_p} + \frac{v_s^2}{2\tilde{t}^2} \right] \qquad (10).$$

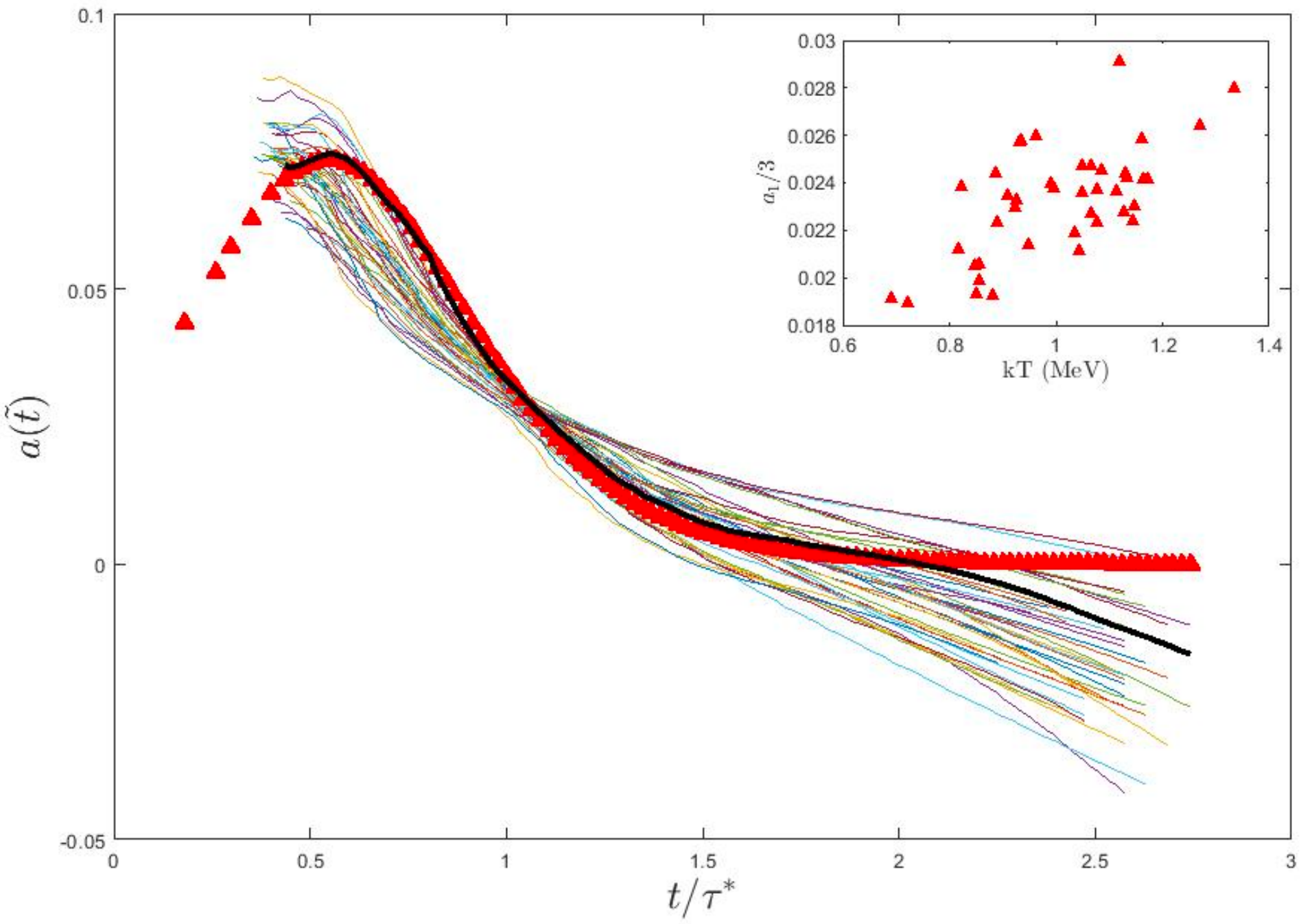

Figure 6. Time evolution of the soliton velocity field, see text. The time is normalized by $\tau^*$, obtained from the condition $\delta E(t=\tau^*) = 0$ for each shot. The thin lines refer to each shot, and (full triangles) the soliton fit to one shot (#429-full line) for illustration, eqs.7,9. (Inset) The parameter $a_1/3$ vs temperature for 41 shots; an almost linear increase is seen as predicted in ref. [24].

In Fig. 6, we have plotted the data corresponding to the right-hand side of eq. (10), thin lines. The data displays a maximum at $\tilde{t} = \widetilde{t_M}$, one of the parameters entering eqs. (7-10). Fitting the free parameter η to the data results in the full triangles in Fig. 6 for a particular shot (#429). The agreement is very good apart at longer times where other phenomena can also play a role. Since the fitting parameters come in the combination $\frac{u}{2}\sqrt{\frac{a_1}{3\beta}}$, it is not possible to further investigate the soliton properties such as its speed. In particular [24], the soliton speed is given by $V_s = u + a_1/3$; the parameter $a_1/3$ can be obtained from eq. 9, but not $u$. $a_1$ increases with increasing temperature from 0.018c to 0.03c (see the inset in Fig. 6).

## V. Comparison to previous experimental results.

In ref. [26], a clever experimental method was introduced to measure the plasma charge deposited in a sensitive detector which measured the integrated net charge as function of time accumulated on it. This charge is the result of positive ions and electrons deposited in the detector. Because of the different masses of the charged particles, they will be deposited at different times, with fast electrons moving essentially at the speed of light, arriving earlier with respect to the ions (contributing to the EMP). At later times, slow electrons may also get deposited in the detector thus reducing the net accumulated positive charge. In ref. [26], the authors performed 3D particles in cells simulations using the CST particle studio code in cylindrical symmetry [27]. They performed numerical simulations by adding to the neutral plasma one proton and one electron with different energies (see figure 4a of ref. [23]). These charged particles propagate in the plasma field and are emitted and revealed in the detector located at distance D (=2.2m) from the target.

To compare our results with ref. [26], we need two simple steps. First transform the time evolution of the plasma from eq. 5 to the time $t_D$ it takes to reach the detector:

$$t_D = t\frac{D}{d/2} = \frac{D}{v} \qquad (11)$$

The second step is to integrate the electric field in time since the net accumulated charge was measured in [26]:

$$E(t) = \frac{1}{\tau^*}\int_{t_0}^{t}\frac{\delta(E)}{D}dt = \frac{1}{D\tau^*}\int_{t_0}^{t}\delta(E)dt. \qquad (12)$$

Where $\delta E(t)/D$ is the electric field produced at time t and accumulated in a time step $\Delta t$, $t_0$ and t are the initial and final times. Here $\delta(E)$ is indeed the difference in energy represented in Fig. 4, produced by charge perturbation. Such charge separation produces a restoring electric potential of the order of $\delta(E)$, and an "infinitesimal" electric field of the order of $\delta(E)/D$.

In Fig. 7, we compare our results (full lines) to ref. [26] (triangles and squares). The qualitative behavior is the same although some differences can be noticed. These differences may be due to the target material used, parylene-N (an insulator) in [26] and Al (a conductor) in our work. The targets thicknesses, 6μm in our work and less than 1μm in [26], and finally the different pulses, laser energies and focalizations at the Vulcan laser facility, see table 1 in ref. [26], and in our work at Vega [1]. In particular, the laser energy at Vulcan is of the order of 350 Joule compared to 7J at Vega III. We stress that our method is based on the detection of energetic protons and the ratios of two nuclear reactions to get the plasma temperature T. In contrast, in ref. [26], the net charge is measured, and we could expect especially protons (and electrons) to be produced at the shortest times. However, charged heavier ions, such as C in different charge states [11,12], are also produced and these may contribute to the net charge especially at longer times.

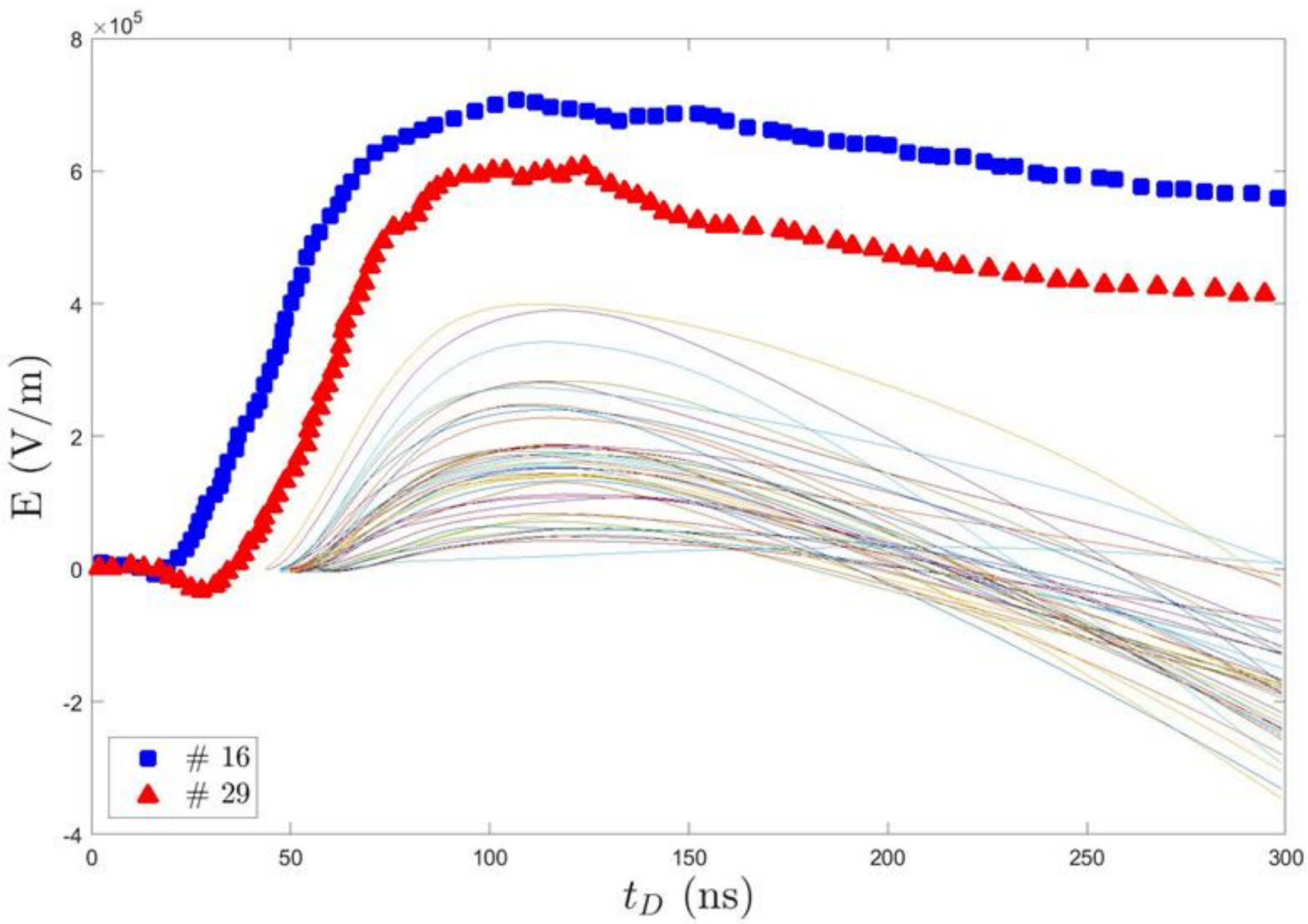

Figure 7. Time evolution of the accumulated net charge in volt/meter. *The full symbols are* from ref. [26], shot #16 (laser energy 348J, pulse duration 1.2ps, target thickness 116 nm) and shot #29 (laser energy 386J, pulse duration 1.7ps, target thickness 269 nm). The lines are our results for 41 shots (laser energy 7J, pulse duration ~200fs, target thickness 6μm). The much larger laser energy at Vulcan is responsible for the higher amplitude of the *electric field.*

Notice that for our shots in Fig. 7 there are no values at short times. These correspond to very energetic protons which were not well measured in our experiment. Thus, we expect some corrections to our results when adding the missing spectrum. As we can see from Fig. 7, we expect such contribution to be small.

We can compare the different results by inverting the procedure outlined above: instead of integrating our $\delta E$, we perform a numerical derivative of the electric field measured in [26]. To get smooth derivatives, we used a fourth-order central difference algorithm based on the Taylor method, where the derivative at a given point is related to the two preceding and the two following points.

In Fig. 8, we compare the results from the two experiments. The qualitative behavior is quite the same and a shift of the peak position is observed when increasing the target thickness, see table 1 in ref. [26]. Since our target was 6μ thick this may be an indication of the different behavior when increasing the target thickness. It would be highly interesting repeating our experiment and the one in ref. [26] by changing the target thickness and material composition while leaving all the laser parameters the same. This may help to optimize the system to get the best performance for the goals we may have, for instance radio nuclide production for medicine, fusion reactions or energetic ion production.

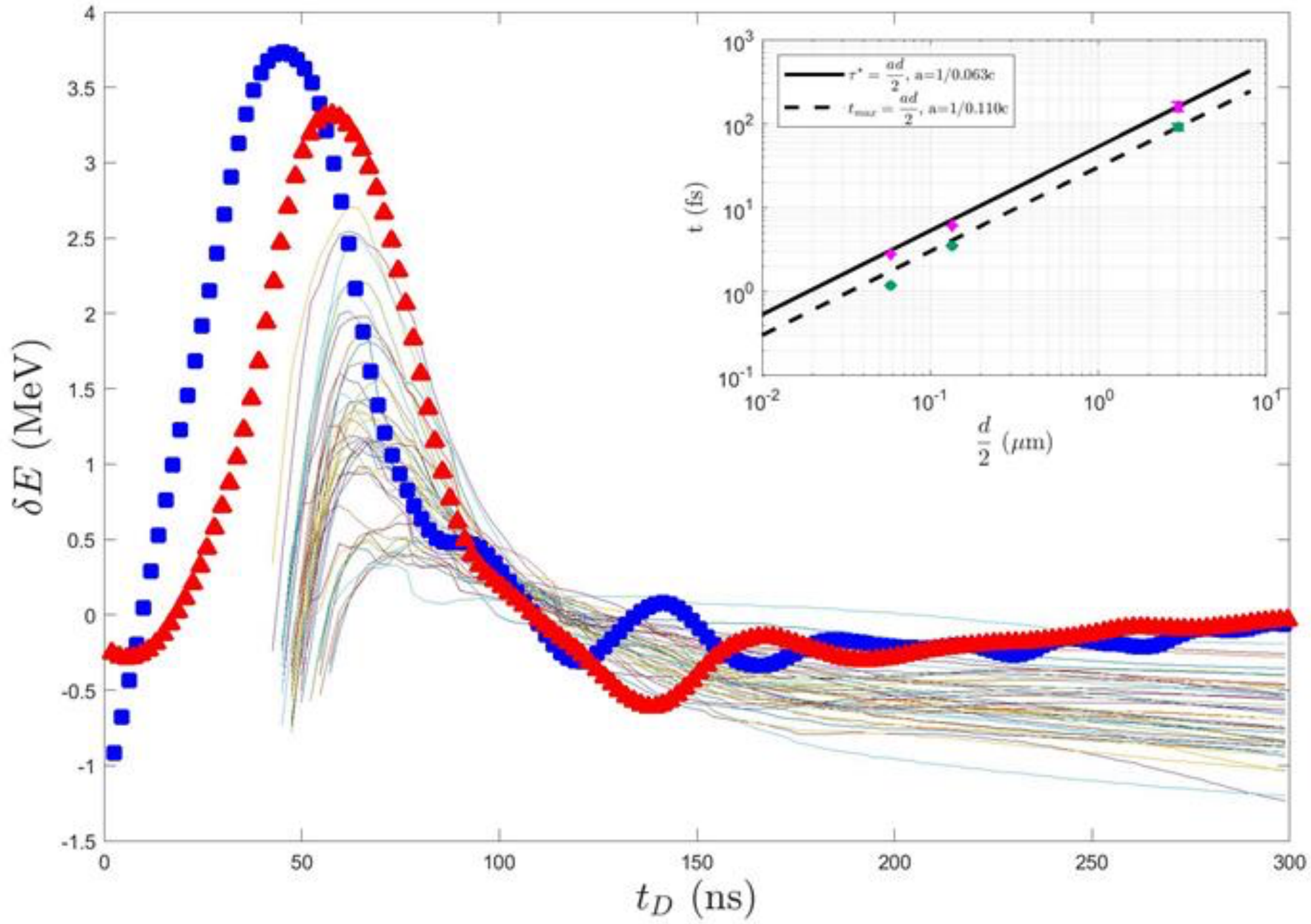

Figure 8. Soliton field electric distribution (in MeV) compared to the time derivative of the electric field measured in ref. [26]. Symbols as in Fig. 7. The numerical time derivative was performed using a fourth order Taylor method. Shots #16 and # 29 from ref. [26] with thickness 116 nm and 269 nm. Notice the shift in the peak position when increasing the thickness for these two shots. Recall that in our experiment a 6μm Al targets were used. In the inset, the time it takes to reach the maximum $t_{max}$ and $\tau^*$ defined before are plotted as function of d/2. Notice that the velocity obtained from the inverse of the fitting constant for $\tau^*$ is very close to the estimate above and plotted in Fig.5. We stress that our results at small times $t_d$ are missing because high energy protons are not properly measured.

In the inset of Fig. 8, we plot the time for the signal to reach the maximum (zero) vs d/2. A good linear relation is obtained and notice that the value of the inverse velocity fitting parameter is very close to the typical velocity $v_s$ derived above. This may be just a coincidence or something that could be used in future experiments to choose the target thickness for a given pulse duration.

## VI Conclusions.

In this paper we have derived nuclear reaction rates using the proton energy distribution for each shot. From the ratio of two well measured isotopes, $^{11}C$ and $^{7}Be$, we have been able to derive an effective temperature reproducing the average radioisotope yield and we extended it to the single shot case. From the knowledge of the effective temperature, and using the ratios of $^{11}C$ and α, as well as $^{7}Be$ and α, we have derived the number of α particles. This is possible since we the reactions cross sections are well known. Finally, the single shot proton energy distribution and temperatures allow us to derive the EoS of the TNSA plasma. Comparison to the ideal gas EoS reveal significant non-equilibrium effects and Coulomb charge dynamics.

Due to charge separation, charge-waves are created in the plasma which generate strong transient fields to restore charge neutrality. We have shown that the behavior of these perturbations to the neutral state of the plasma is consistent with the Korteweg-de Vries equation valid for a quasi-neutral plasma. The KdV equation describes the soliton propagation in the plasma with sub luminal speeds. The plasma non-neutrality lasts for times less than the laser pulse duration and depends on the target thickness. We have reanalyzed the experiment of ref. [26] and demonstrate its consistency to our results. Future experiments should be dedicated to systematic studies of the soliton dynamics when changing the target thickness and composition, the laser pulse duration, energy and focalization. Understanding the soliton dynamics is crucial for applications and basic science and may be extended to astrophysical scenarios.

## Author Contributions

AB devised the scheme to study the fluctuations, wrote the first draft of paper, and co-wrote the final version of the paper. MRDR performed the data analysis of the HPGE. ME, EF and DR did the data analysis of the zero-degree TS. ZZ analyzed the data, performed the numerical calculations and prepared all the figures. DB was the spokesperson of the March 2023 experiment and co-wrote the final version of the paper. All authors contributed to the experiment proposal, preparation and the final form of the manuscript. Inspired by the results of this work, a theoretical paper was recently published [28] incorporating the general aspects of the soliton dynamics.

## Funding

This work was supported in part by the United States Department of Energy under Grant #DE-FG02-93ER40773; the EUROfusion Consortium, funded by the European Union via the Euratom Research and Training Programme (Grant Agreement No 101052200 - EUROfusion); the COST Action CA21128 - PROBONO "PROton Boron Nuclear fusion: from energy production to medical applicatiOns", supported by COST (European Cooperation in Science and Technology - www.cost.eu). We also acknowledge the Unidad de Investigación Consolidada (UIC 167), Junta de Castilla y León Grant No. CLP087U16. ZZ was supported in part by the University of Chinese Academy of Sciences, Beijing 100049, China. ZZ and JX received support from the National Natural Science Foundation of China under Grant No.12375125, and the Fundamental Research Funds for the Central Universities.

## Acknowledgments

A particular thank you to all the personnel at the CLPU-Vega facility for their direct and indirect help. We thank Mr. A. Massara (LNS-INFN, Catania-Italy) for providing to specification the thin targets used in the experiment. We thank Dr. B. Roeder for the digital acquisition support.

**Competing interests:** The authors declare no competing interests

## Appendix

A schematic layout of the experiment performed at Vega III is given in Fig. 9.

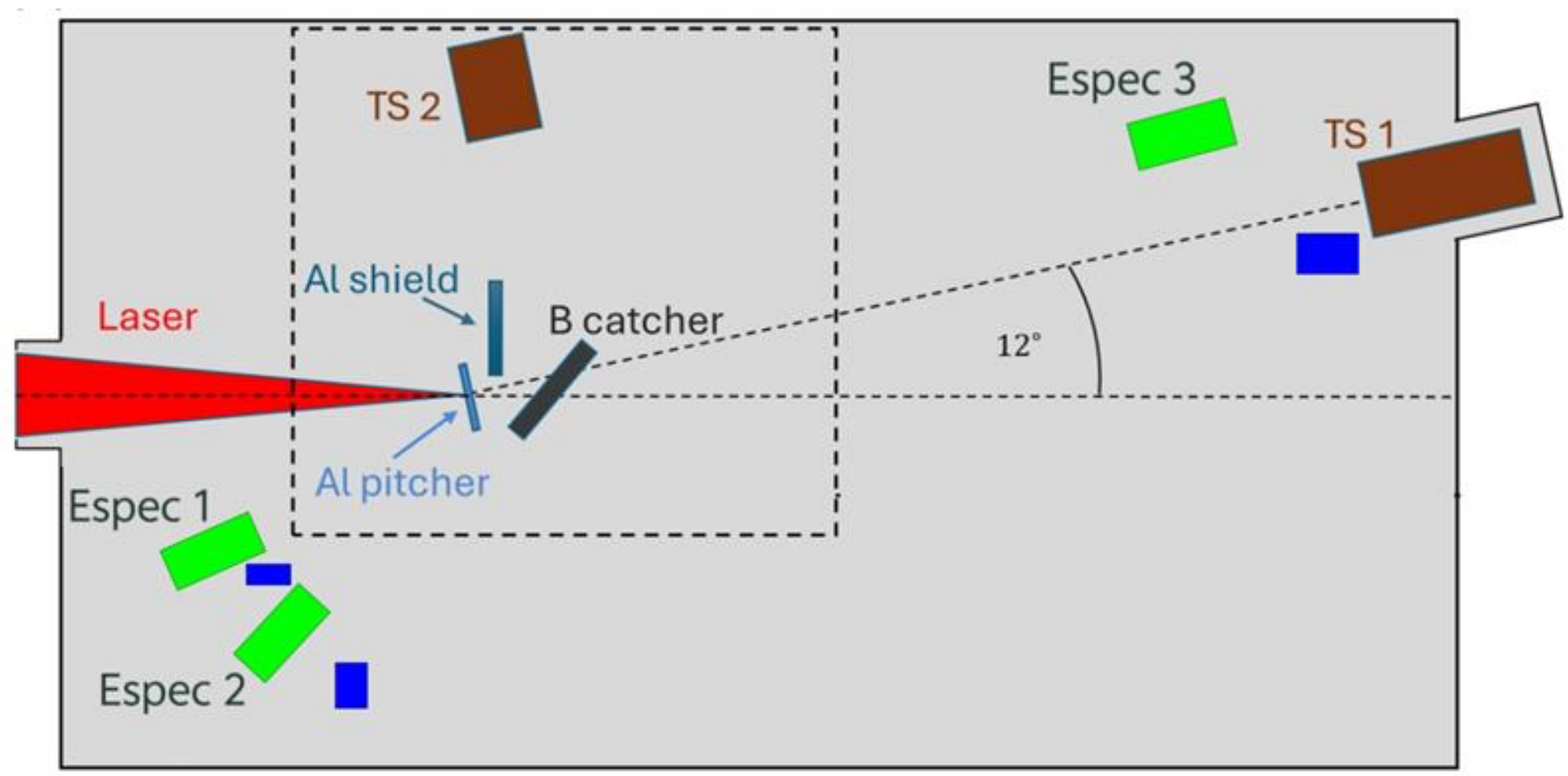

Figure 9. Schematic view of the experimental setup. The laser is impinging on the Al target slightly tilted respect to the laser direction to avoid dangerous reflections which could damage the optics. A B catcher is facing the Al about 2 cm away from it. The catcher can be removed for some shots. When there is no B, the protons and heavier ions are produced perpendicular to the Al target and detected by the TS1 (this defines zero degrees [1]). Cu targets used as catcher are located at different angles; some are indicated by the blue squares. Three more spectrometers for electrons were located as indicated in the figure. Electron spectrometer Espec1 and 3 can measure electrons up to 25MeV, while Espec2 measures up to 5MeV. A second TS2 for low energy protons (up to about 350KeV) was located at about 90 degrees from TS1 and it was used to measured electrons scattered from the B catcher

In this appendix we are going to discuss the energy distributions obtained from electron spectrometers and located at different angles as displayed in Fig. 9. Also in this case, IP were used to record the electron signals which were accumulated over a certain number of shots, as in the proton case. In Fig. 10 we compare the proton and electron distributions. They all show a high energy tail, except the one recorder with the electron spectrometer Espec2 which by construction can detect electrons only up to 5MeV. Electrons of energy of the order of 0.5MeV and above practically move at the speed of light while the protons of similar energies are non-relativistic and practically at rest with respect to them. Thus, the electrons leave the plasma which becomes positively charged and the residual produced Coulomb potential accelerates the protons and heavier ions. Recall that the protons and heavier ions (C and O mainly) are just coming from impurities on the surface the Al target. Thus, the number of accelerated particles is small in comparison to the number of Al ions and electrons. In these conditions, the plasma is quasi neutral and the charged protons moving inside it give rise to solitons as first discussed by Lifshitz and Pitaevskii [20] more than 60 years ago. Electron distributions were measured above 9 degrees respect to TS1 (Espec3) while other two e-spectrometers were located in the back direction. Proton spectra are very strongly peaked in the forward direction and their yield at similar angles of the electron spectrometer are estimated to be more than 3 orders of magnitude less than the yields at zero degrees [1]. This is also confirmed by the much lower Cu activation at large angles, as shown in Fig.7 of ref. [1]. Hence, we expect the electron spectra to be very similar to the proton ones at the same angles, a feature which would be very interesting to test in order to validate theoretical models. In future experiments it would be wise to exchange the electron spectrometer with the proton one, for instance Espec3 with TS1, as well as Espec1 with TS1. For completeness, in Fig.10 we have added the protons measured by the TS2 detector even though it was utilized for different shots when the B-catcher was included [11,12]. Thus, it is sensitive to protons which have been backscattered by the B catcher showing a yield close to the low energy electrons measured at different angles. A final comment for the spectra measured by electron spectrometers 1 and 3. They are located in opposite directions respect to the Al target and show a remarkable similar yield up to about 2 MeV. However, the electrons in the forward direction display a much higher yield at larger energies.

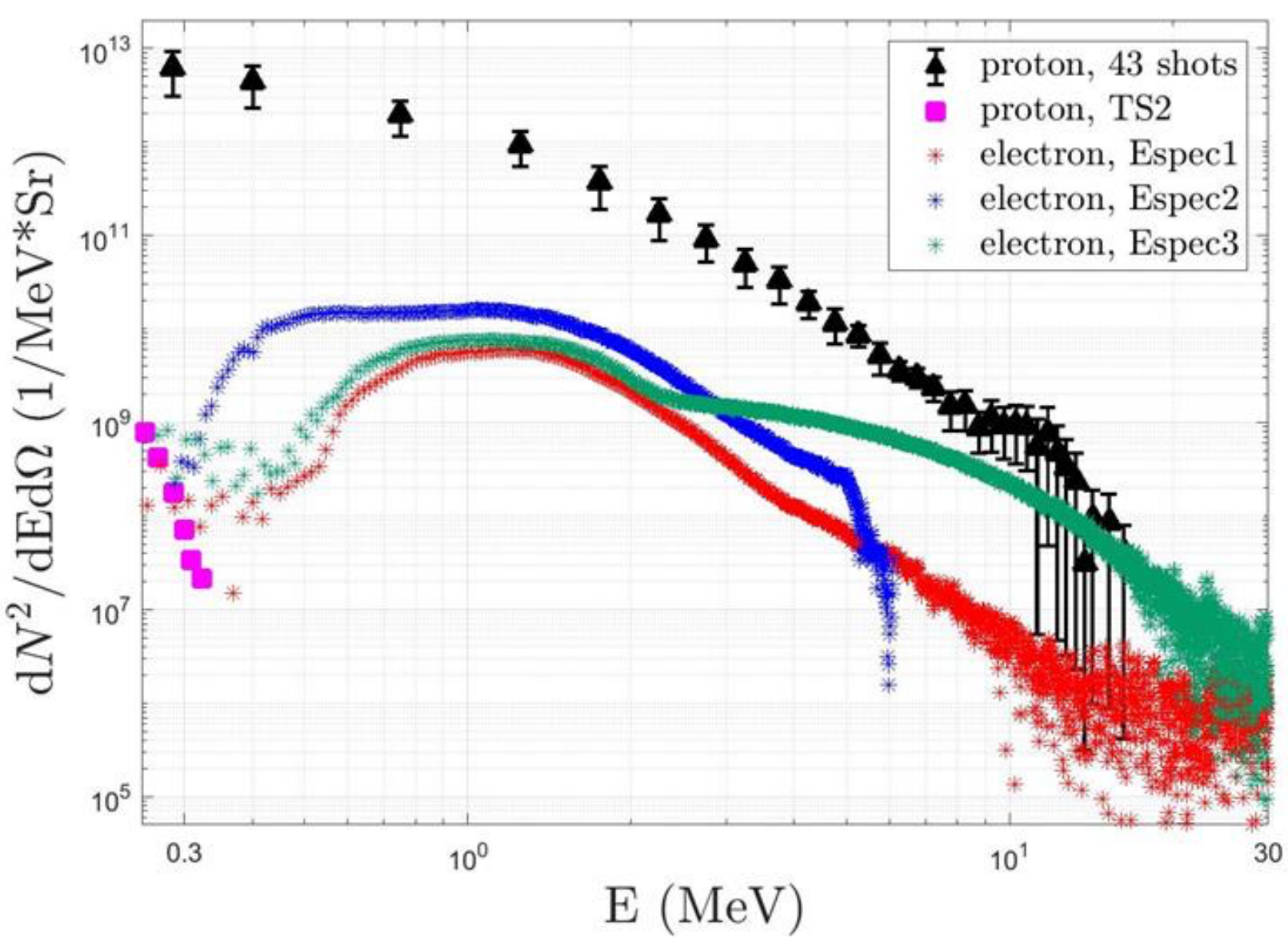

Figure 10. Proton and electron distributions averaged over 43 and 80 shots respectively. The low energy proton spectrometer (TS2) was averaged over 71 shots.

For completeness, in Fig. 11 we plot the detected carbon yields for each charge state [12]. We notice that in ref. [26], the total charge (i.e. proton. C and other ions) is integrated over time while our analysis is for protons only. This may explain the differences observed in Figs.8 and 9 especially at longer times probably due to heavier (positively charged) ions moving slower because of their larger masses.

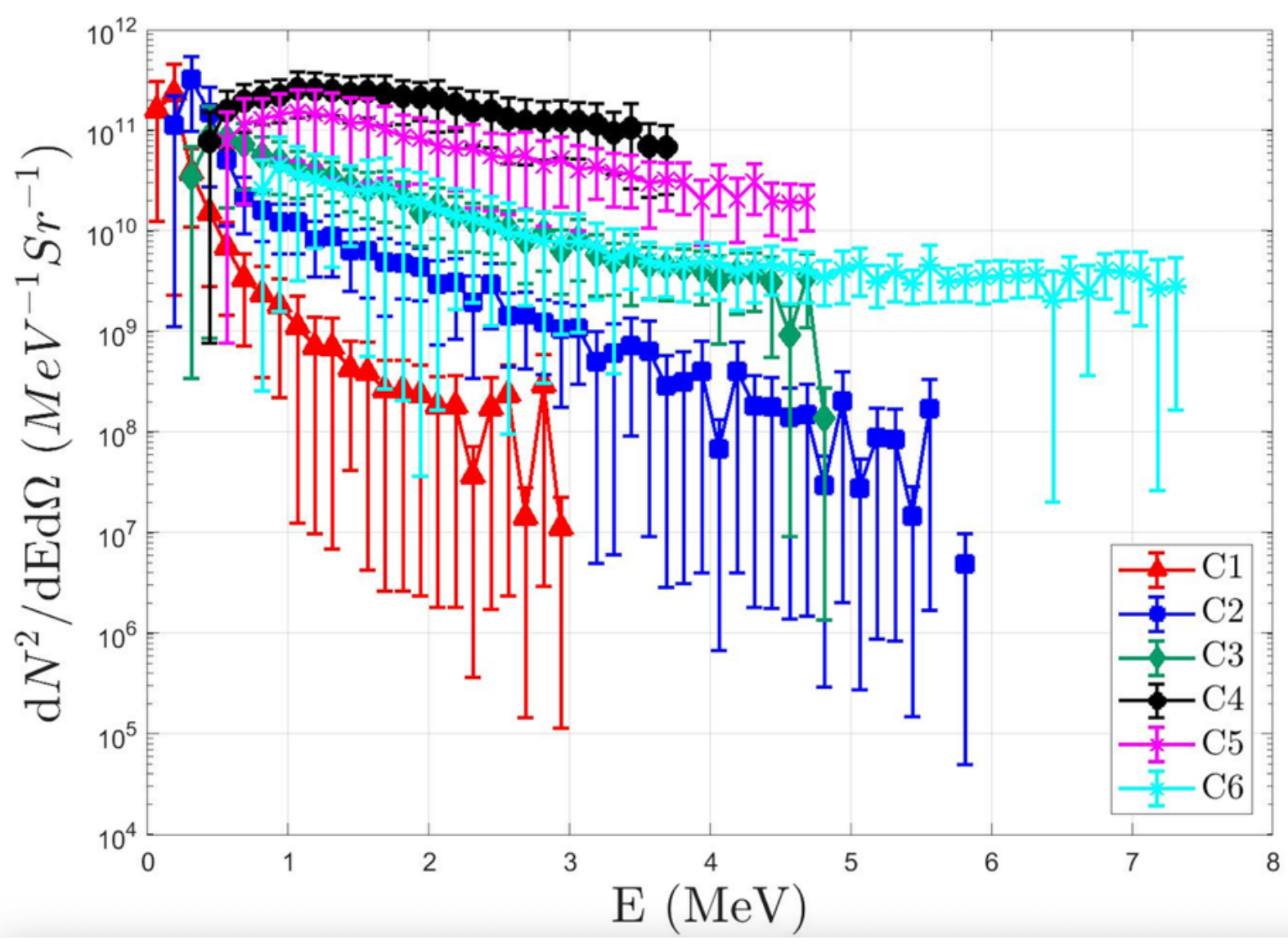

Figure 11. Same as Fig. 10 but for different charge states of C ions as indicated in the inset.